\documentclass[aps,prc,preprint,groupedaddress,showpacs]{revtex4-1}

\usepackage{graphicx}
\usepackage{amssymb}

\begin{document}


\title{Predictions of hadronic observables in $Pb+Pb$ collisions at $\sqrt {s_{NN}}=2.76$ TeV
from a hadronic rescattering model}


\author{T. J. Humanic}
\email[]{humanic@mps.ohio-state.edu}
\affiliation{Department of Physics, The Ohio State University,
Columbus, Ohio, USA}


\date{\today}

\begin{abstract}
A kinematic model based on the superposition of $p+p$ collisions, relativistic geometry and final-state hadronic rescattering
is used to predict various hadronic observables in $\sqrt {s_{NN}} = 2.76$ TeV $Pb+Pb$ collisions. 
Predictions for rapidity and transverse momentum distributions, elliptic flow, and two-boson femtoscopy are presented. A short proper time for hadronization is assumed. Previous calculations using this
model which were performed
for $\sqrt {s_{NN}}=200$ GeV $Au+Au$ collisions  were shown to describe reasonably well the trends of
observables in experiments carried out at that energy, giving the present predictions for $Pb+Pb$
at higher energy some degree of credibility.
\end{abstract}


\maketitle

\section{Introduction\label{introduction}}
The present plan for running at the CERN Large Hadron Collider (LHC) calls
for $Pb+Pb$ collisions at $\sqrt{s_{NN}}=$ 2.76 TeV to commence in November, 2010.
These will be the highest energy heavy ion collisions ever produced in the laboratory, and
it is hoped that exotic phenomena will be observed which will expand our knowledge of 
the properties of highly excited matter \cite{Aamodt:2008zz}.

In this context, it is the goal of the present paper to make predictions for common
hadronic observables which will be initially measured by LHC experiments in these
high-energy $Pb+Pb$ collisions. Predictions for spectra (i.e. rapidity and transverse momentum distributions), 
elliptic flow, and two-boson femtoscopy are presented. A hadronic rescattering model
in which the initial state is determined by the superposition of proton-proton collisions
has been chosen to make these predictions. The advantages of this 
model for this purpose are 1) the model
has been shown to describe the overall trends of hadronic observables in lower energy
$Au+Au$ collisions at $\sqrt{s_{NN}}=$ 0.20 TeV from the Relativistic Heavy Ion Collider 
(RHIC)\cite{Humanic:2008nt}, and 2) the model is easily scalable to LHC energies.
These will be ``limiting case scenario'' predictions in the sense that only hadrons are
used as the degrees of freedom in this model even at the early stages of the
collision where parton degrees of freedom are thought to be more appropriate, i.e. a short
proper time for hadronization is assumed.

The paper is organized into the following sections: Section \ref{description} gives a brief description of the model, Section \ref{predictions} presents predictions from the model for
$\sqrt{s_{NN}}=2.76$ TeV $Pb+Pb$ collisions, and Section \ref{conclusions} gives a summary and conclusions.

\section{Description of the model\label{description}}
The model calculations are carried out in five main steps: A) generate hadrons in $p+p$  collisions from PYTHIA, B) superpose $p+p$ collisions in the geometry of the colliding nuclei, C) employ a simple space-time geometry picture for the hadronization of the
PYTHIA-generated hadrons,   D) calculate the effects of final-state rescattering among the hadrons,
and E) calculate the hadronic observables. These steps will now be discussed in more detail.

\subsection{Generation of the $p+p$ collisions with PYTHIA}
The $p+p$ collisions were modeled with the PYTHIA code \cite{pythia6.4}, version 6.409. The internal parton distribution functions ``CTEQ 5L'' (leading order) were used in these calculations. Events were generated
in ``minimum bias'' mode, i.e. setting the low-$p_T$ cutoff for parton-parton collisions to zero (or
in terms of the actual PYTHIA parameter, $ckin(3)=0$) and excluding elastic and diffractive collisions (PYTHIA parameter $msel=1$). Runs were made at $\sqrt{s}= $ 2.76 TeV to simulate 
the upcoming LHC collisions. Information saved
from a PYTHIA run for use in the next step of the procedure were the momenta and identities
of the ``direct'' (i.e. redundancies removed) hadrons (all charge states) $\pi$, $K$, $p$, $n$,
$\Delta$, $\Lambda$, $\rho$, $\omega$, $\eta$, ${\eta}'$, $\phi$, and $K^*$. These particles were
chosen since they are the most common hadrons produced and thus should have the greatest
effect on the hadronic observables in these calculations.

\subsection{Superposition of $p+p$ events to simulate heavy-ion collisions}
An assumption of the model is that an adequate job of describing the heavy-ion collision can be obtained by superposing PYTHIA-generated  $p+p$ collisions calculated at the beam $\sqrt s$
within the collision geometry of the colliding nuclei. Specifically, for a collision of impact
parameter $b$, if $f(b)$ is the fraction of the overlap volume of the participating parts of the nuclei such that $f(b=0)=1$ and $f(b=2R)=0$, where $R=1.2A^{1/3}$ and $A$ is the mass 
number of the nuclei, then
the number of $p+p$ collisions to be superposed will be $f(b)A$. The positions of the superposed $p+p$ pairs are randomly distributed in the overlap volume and then projected onto the $x-y$ plane which is transverse to the beam axis defined in the $z$-direction. The coordinates for a particular
$p+p$ pair are defined as $x_{pp}$, $y_{pp}$, and $z_{pp} = 0$. 
The positions of the hadrons produced in 
one of these $p+p$ collisions are defined with respect to the position so obtained of the superposed
$p+p$ collision (see below). 

As was done in similar calculations for RHIC collisions to give better 
agreement with experimental $dn/d\eta$ distributions\cite{Humanic:2008nt}, a lower multiplicity cut
was applied to the $p+p$ collisions used in the present calculations which 
rejected the lowest ~20\% of the events. The spirit of this cut is to partially compensate for the fact that there is no reinteraction of primary nucleons from the projectile-target system in this model.

\subsection{The space-time geometry picture for hadronization}
The space-time geometry picture for hadronization from a superposed $p+p$
collision located at $(x_{pp},y_{pp})$ consists of the emission of a PYTHIA
particle from a thin uniform disk of radius 1 fm in the $x-y$ plane followed by
its hadronization which occurs in the proper time of the particle, $\tau$. The space-time
coordinates at hadronization in the lab frame $(x_h, y_h, z_h, t_h)$ for a particle with momentum
coordinates $(p_x, p_y, p_z)$, energy $E$, rest mass $m_0$, and transverse disk
coordinates $(x_0, y_0)$, which are chosen randomly on the disk,  can then be written as

\begin{eqnarray}
x_h = x_{pp} + x_0 + \tau \frac{p_x}{m_0} \\
y_h = y_{pp} + y_0 + \tau \frac{p_y}{m_0} \\
z_h = \tau \frac{p_z}{m_0} \\
t_h = \tau \frac{E}{m_0}
\end{eqnarray}

Eqs. (1) and (2) show the initial expansion in the transverse direction now present in the model. 
The simplicity of this geometric picture is now clear: it is just an expression of causality with the
assumption that all particles hadronize with the same proper time, $\tau$. A similar hadronization
picture (with an initial point source) has been applied to $e^+-e^-$ collisions\cite{csorgo}.
For all results presented in this work,  $\tau$ will be set to 0.1 fm/c as was done in
applying the present model to calculating predictions for 
RHIC $Au+Au$ collisions\cite{Humanic:2008nt} and 
Tevatron $p+\bar{p}$ collisions\cite{Humanic:2006ib}.

\subsection{Final-state hadronic rescattering}
The hadronic rescattering calculational method used is similar to that
employed in previous studies \cite{Humanic:1998a,Humanic:2006a}.
Rescattering is simulated with a semi-classical Monte Carlo
calculation which assumes strong binary collisions between hadrons.
Relativistic kinematics is used throughout. The hadrons considered in the
calculation are the most common ones: pions, kaons,
nucleons and lambdas ($\pi$, K,
N, and $\Lambda$), and the $\rho$, $\omega$, $\eta$, ${\eta}'$,
$\phi$, $\Delta$, and $K^*$ resonances.
For simplicity, the
calculation is isospin averaged (e.g. no distinction is made among a
$\pi^{+}$, $\pi^0$, and $\pi^{-}$).

The rescattering calculation finishes
with the freeze out and decay of all particles. Starting from the
initial stage ($t=0$ fm/c), the positions of all particles in each event are
allowed to evolve in time in small time steps ($\Delta t=0.5$ fm/c)
according to their initial momenta. At each time step each particle
is checked to see a) if it has hadronized ($t>t_h$, where $t_h$ is given in
Eq. (4)), b) if it
decays, and c) if it is sufficiently close to another particle to
scatter with it. Isospin-averaged s-wave and p-wave cross sections
for meson scattering are obtained from Prakash et al.\cite{Prakash:1993a}
and other cross sections are estimated from fits to hadron scattering data
in the Review of Particle Physics\cite{pdg}. Both elastic and inelastic collisions are
included. The calculation is carried out to 400 fm/c which
allows enough time for the rescattering to finish (as a test, calculations were also carried out for
longer times with no changes in the results). Note that when this cutoff time is reached, all un-decayed resonances are allowed to decay with their natural lifetimes and their projected decay positions and times are recorded.

The rescattering calculation is described in more detail elsewhere
\cite{Humanic:2006a,Humanic:1998a}. The validity of the numerical
methods used in the rescattering code have been studied using
the subdivision method, the results of which have verified that the methods used are 
valid \cite{Humanic:2006b}.

\subsection{Calculation of the hadronic observables}
Model runs are made to be ``minimum bias'' by having the impact parameters of collisions follow the distribution $d\sigma/db \propto  b$, where $0<b<2R$. Observables are then calculated from the model in the appropriate centrality bin by making multiplicity cuts as normally done in experiments, as well as kinematic cuts on rapidity and $p_T$. For the present study, a full-calculation 3200 event minimum bias run was made from the model for $\sqrt {s_{NN}}=2.76$ TeV $Pb+Pb$ collisions which was then used to calculate all of the hadronic observables shown. In addition, a 3200 event minimum bias
run with rescattering turned off in the model was also made for comparison to study the
importance of rescattering in these observables.

\section{Predictions for $\sqrt{s_{NN}}=$ 2.76 TeV $Pb+Pb$ collisions\label{predictions}}
As mentioned earlier, predictions for the hadronic observables spectra, elliptic flow, and
two-boson femtoscopy have been made with the present model 
for $\sqrt{s_{NN}}=$ 2.76 TeV $Pb+Pb$ collisions. Results for each of these observables
are presented separately below.

\subsection{Spectra}
\begin{figure}
\begin{center}
\includegraphics[width=100mm]{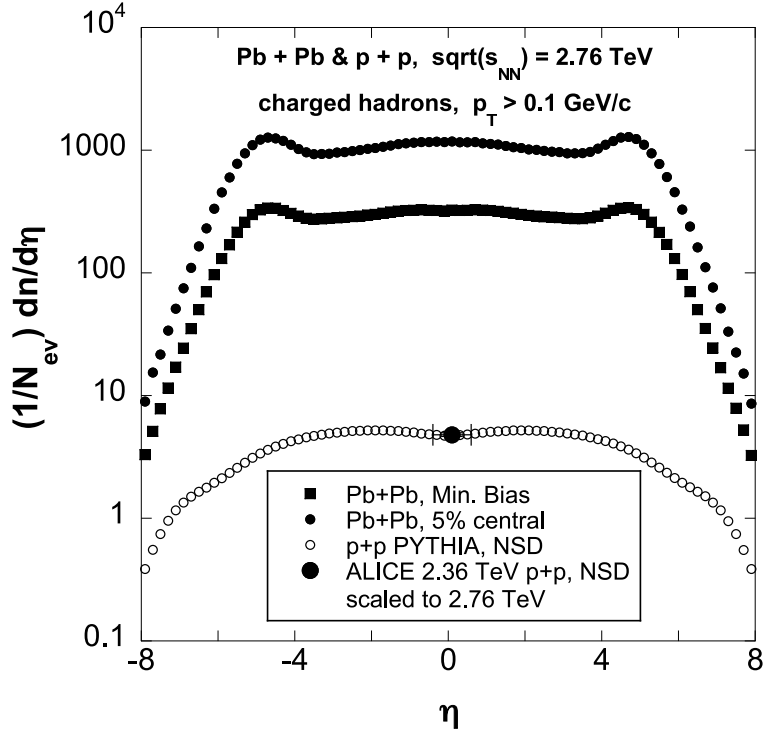} \caption{Model charged-hadron pseudorapidity distributions for $Pb+Pb$ collisions at $\sqrt {s_{NN}} = 2.76$ TeV for minimum bias and $0-5\%$ centrality bins.
Also shown is a comparison of non-single diffractive PYTHIA $p+p$ collisions at $\sqrt {s} = 2.76$ TeV
with a measurement from ALICE scaled to this energy. }
\label{fig1}
\end{center}
\end{figure}

\begin{figure}
\begin{center}
\includegraphics[width=100mm]{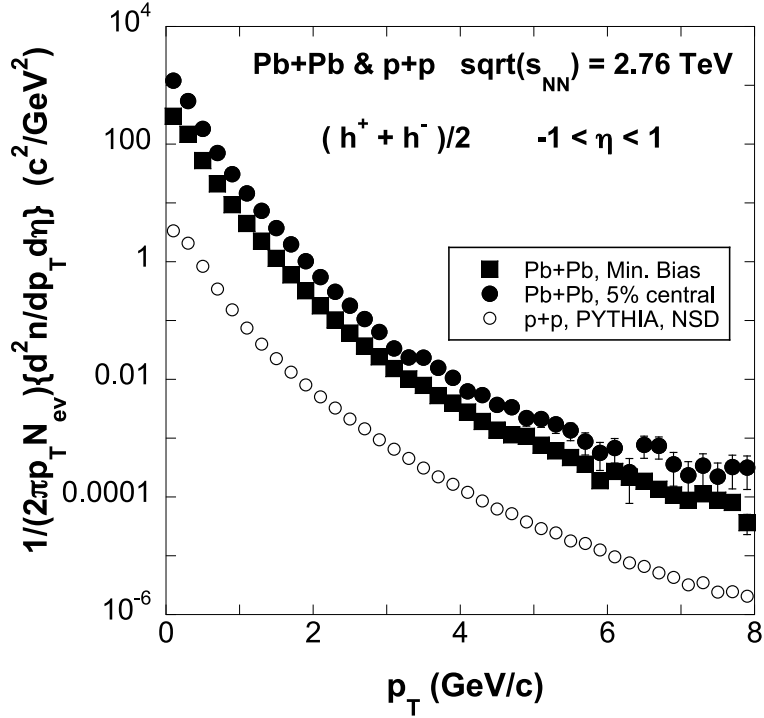} \caption{Model charged-hadron $p_T$ distributions
for the same conditions as shown in Figure \ref{fig1}.}
\label{fig2}
\end{center}
\end{figure}

\begin{figure}
\begin{center}
\includegraphics[width=140mm]{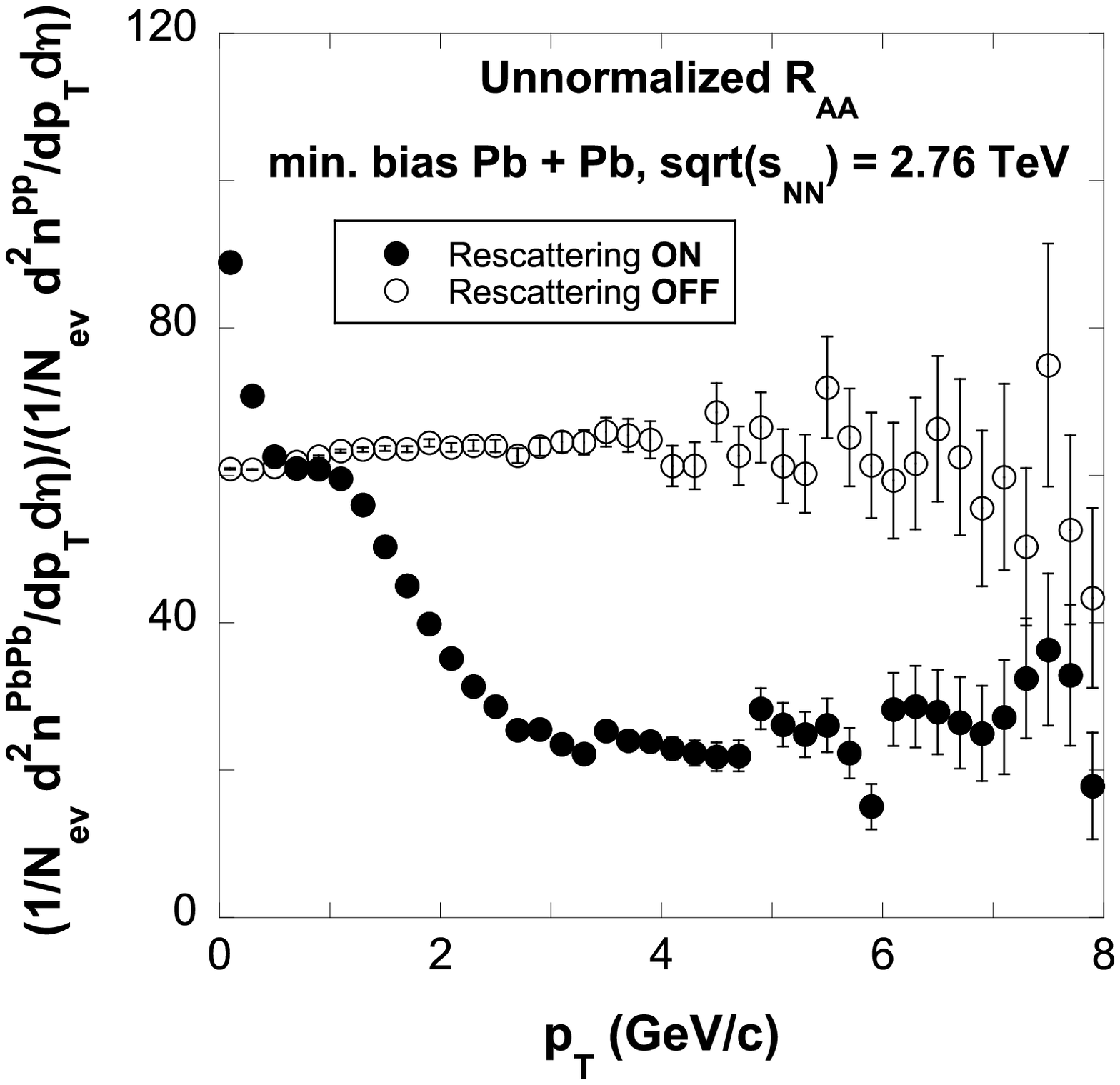} \caption{Ratio of minimum bias $Pb+Pb$ to
$p+p$ $p_T$ distributions from Figure \ref{fig2} (indicated as ``Rescattering ON''). Also shown is
the similar ratio with rescattering turned off in the model.}
\label{fig3}
\end{center}
\end{figure}

Figures \ref{fig1}, \ref{fig2}, and \ref{fig3} show predictions from the present model for charged hadron spectra.

Figure \ref{fig1} shows charged-hadron pseudorapidity distributions for $Pb+Pb$ collisions at $\sqrt {s_{NN}} = 2.76$ TeV for minimum bias and $0-5\%$ centrality bins.
Also shown is a comparison of non-single diffractive PYTHIA $p+p$ collisions at $\sqrt {s} = 2.76$ TeV
with a measurement of non-single-diffractive $dn/d\eta$ from the LHC ALICE experiment made
at $\sqrt{s}=2.36$ TeV \cite{Aamodt:2010ft}, approximately scaled to the slightly higher energy
by $\sqrt{2.76/2.36}$. Since the model is ``isospin averaged'', the model distributions are multiplied by $2/3$ to approximate all charged particles. Looking at the $p+p$ distribution first, it is seen that
PYTHIA is in good agreement with the scaled ALICE measurement at mid-rapidity, giving
additional confidence in using PYTHIA $p+p$ collisions at this energy in the superposition
for the $Pb+Pb$ collisions. Looking at the distributions for $Pb+Pb$, the mid-rapidity
$dn/d\eta$ values for minimum bias and central collisions are found to be 323 and 1174, 
respectively. For central collisions, this is about twice the value found in RHIC
$\sqrt{s_{NN}}=200$ GeV $Au+Au$ collisions \cite{phobos1}, and it is at the lower 
end of the range of predictions which have been recently made of $1500-4000$ in central collisions using various extrapolations of RHIC experimental rapidity densities \cite{Aamodt:2008zz}.

Figure \ref{fig2} shows charged-hadron $p_T$ distributions for $Pb+Pb$ collisions at $\sqrt {s_{NN}} = 2.76$ TeV for minimum bias and $0-5\%$ centrality bins.
Also shown for comparison is the $p_T$ distribution for non-single 
diffractive PYTHIA $p+p$ collisions at $\sqrt {s} = 2.76$ TeV.
To approximate $(h^++h^-)/2$ for the model, the model distributions are multiplied by $1/3$.
The $0-5\%$ centrality charged particle $p_T$ distribution for $p_T>5$ GeV/c is predicted to be about two orders of magnitude larger at the LHC compared with RHIC \cite{rhic4}. This is
an expected consequence of the higher $\sqrt {s_{NN}}$ in the 
LHC collisions that the $p_T$ distributions
at high $p_T$ should be greatly enhanced.

Figure \ref{fig3} shows the ratio of minimum bias $Pb+Pb$ to
$p+p$ $p_T$ distributions from Figure \ref{fig2} (indicated as ``Rescattering ON''). Also shown
for comparison is
the similar ratio with rescattering turned off in the model. This ratio is essentially the
``unnormalized'' $R_{AA}$ observable\cite{Adler:2003au}. 
Studying the high $p_T$ behavior of $R_{AA}$
is thought to be a way of more directly studying QCD processes, such as jets, 
in heavy-ion collisions. Since the present model is based on
using PYTHIA which uses QCD processes in calculating $p+p$ collisions, the model
should contain these effects and thus should be suitable for comparing with experiments
which measure these observables. As seen in Figure \ref{fig3}, for the full rescattering calculation
the ratio is suppressed for $p_T>2$ GeV/c compared with the case of rescattering turned off.
This high-$p_T$ suppression is similar to that observed in RHIC collisions\cite{Adler:2003au}.
In the present model calculations this suppression is clearly seen to be due to hadronic
rescattering.



%


\subsection{Elliptic flow}
The elliptic flow variable, $V_2$, is
defined as
\begin{eqnarray}
\label{v2} V_2=<\cos(2\phi)> \\\nonumber
    \phi=\arctan(\frac{p_y}{p_x})
\end{eqnarray}
where ``$<>$'' implies a sum over particles in an event and a sum over events
and where $p_x$ and $p_y$ are the $x$ and $y$ components of the particle
momentum, and $x$ is in the impact parameter direction, i.e.
reaction plane direction, and $y$ is in the direction perpendicular
to the reaction plane. The $V_2$ variable is calculated from the model
using Eq. (\ref{v2}) and taking the reaction plane to be the model $x-z$ plane.

\begin{figure}
\begin{center}
\includegraphics[width=100mm]{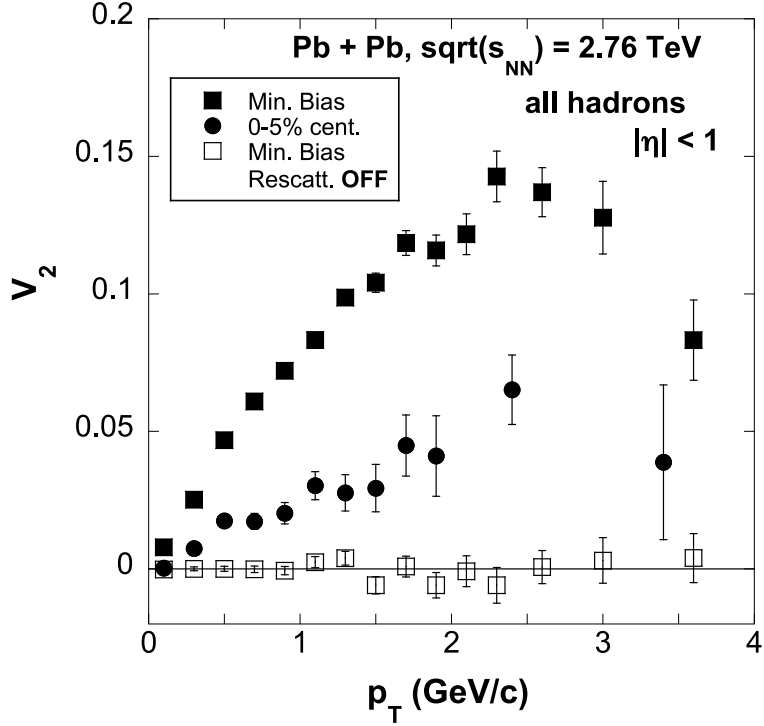} \caption{Model $V_2$ vs. $p_T$ plots for $Pb+Pb$
collisions for minimum bias and $0-5\%$ centrality bins, all hadrons, and $-1<\eta<1$.
Also shown is the minimum bias case with rescattering turned off.}
\label{fig4}
\end{center}
\end{figure}

Figure \ref{fig4} shows model $V_2$ vs. $p_T$ plots for $\sqrt{s_{NN}}=2.76$ TeV $Pb+Pb$
collisions for minimum bias and $0-5\%$ centrality bins, all hadrons, and $-1<\eta<1$.
Also shown is the minimum bias case with rescattering turned off. For the
minimum bias case with the full calculation, $V_2$ is seen to increase with increasing $p_T$, peaking at
a value of about 0.14 and then decreasing for $p_T$ increasing beyond 2.5 GeV/c.
For the $0-5\%$ centrality case, $V_2$ is also seen to increase with increasing $p_T$ and
to have a significantly smaller magnitude than the minimum bias case. These general behaviors
of $V_2$ are also observed in RHIC collisions\cite{Adams:2004bi,Ab:2008ed}. When
rescattering is turned off, $V_2\rightarrow0$ for all $p_T$, demonstrating that hadronic rescattering
accounts completely for the elliptic flow signal in this model.

\subsection{Two-boson femtoscopy (Hanbury-Brown-Twiss interferometry)}
\begin{figure}
\begin{center}
\includegraphics[width=100mm]{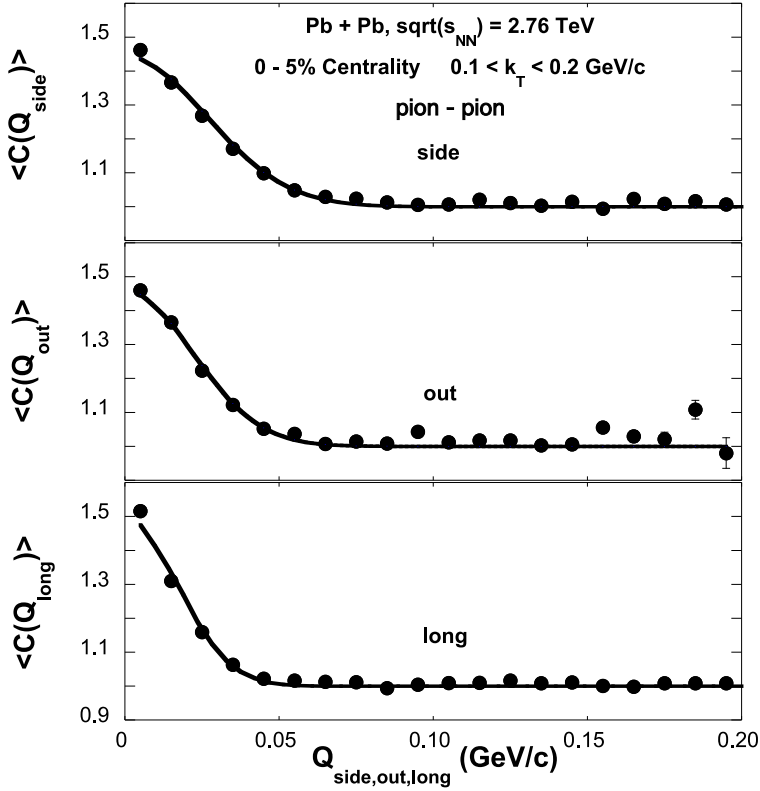} \caption{Sample two-pion correlation
function with Gaussian fit projected onto the $Q_{out}$, $Q_{side}$, and $Q_{long}$
axes from the model. The collision centrality is $0-5\%$ and the $k_T$ bin is $0.1-0.2$ GeV/c.}
\label{fig5}
\end{center}
\end{figure}

\begin{figure}
\begin{center}
\includegraphics[width=140mm]{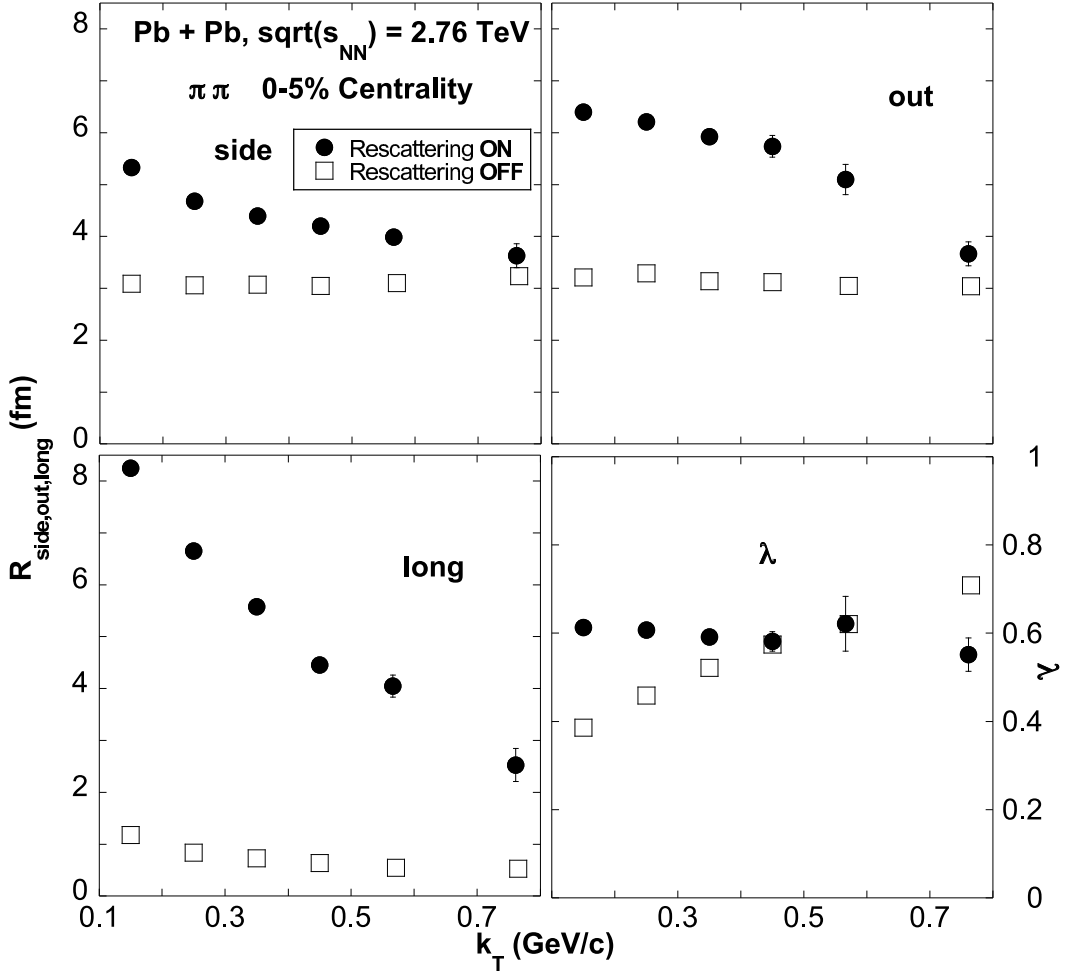} \caption{Model pion source parameters vs. $k_T$ 
for $Pb+Pb$ collisions, $0-5\%$ centrality, $-1<\eta<1$, and with and without rescattering turned
on.}
\label{fig6}
\end{center}
\end{figure}

\begin{figure}
\begin{center}
\includegraphics[width=140mm]{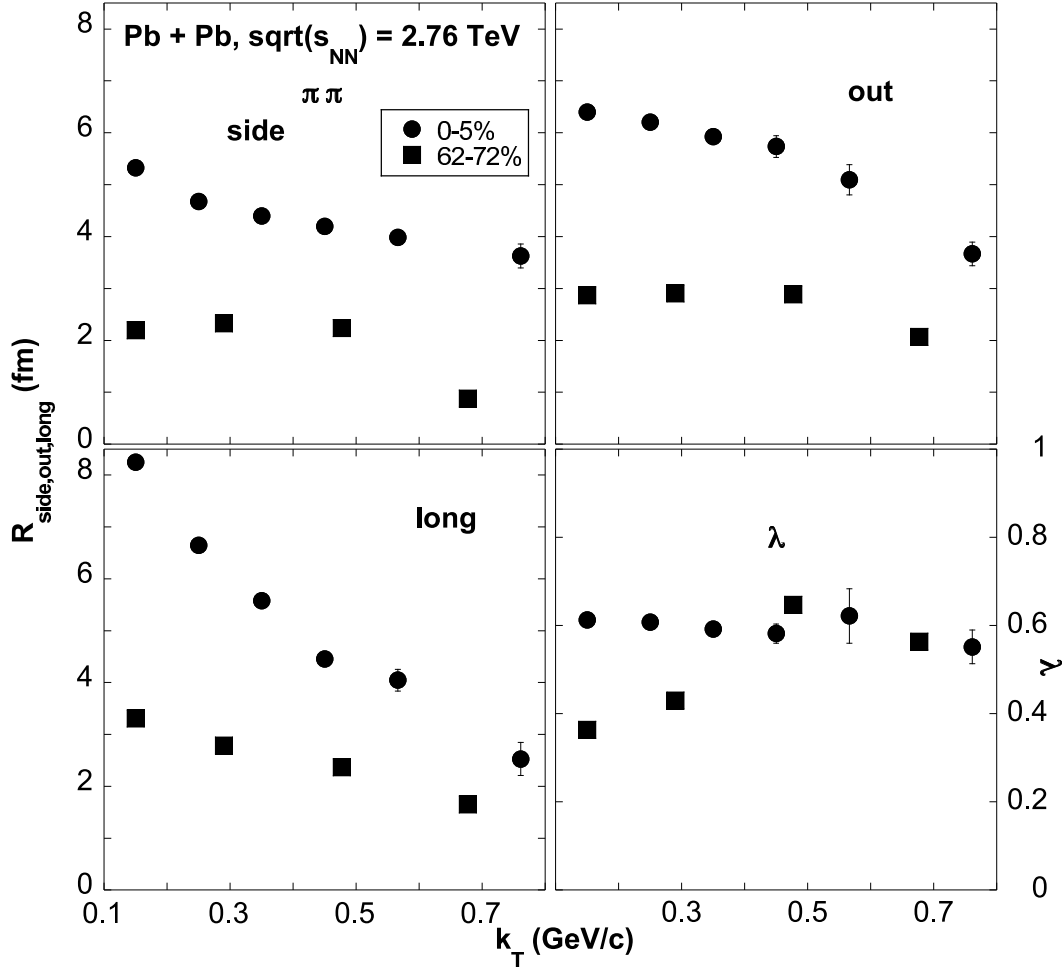} \caption{Model pion source parameters vs. $k_T$ 
for $Pb+Pb$ collisions, $-1<\eta<1$, $0-5\%$ and $62-72\%$ centralities.}
\label{fig7}
\end{center}
\end{figure}

\begin{figure}
\begin{center}
\includegraphics[width=100mm]{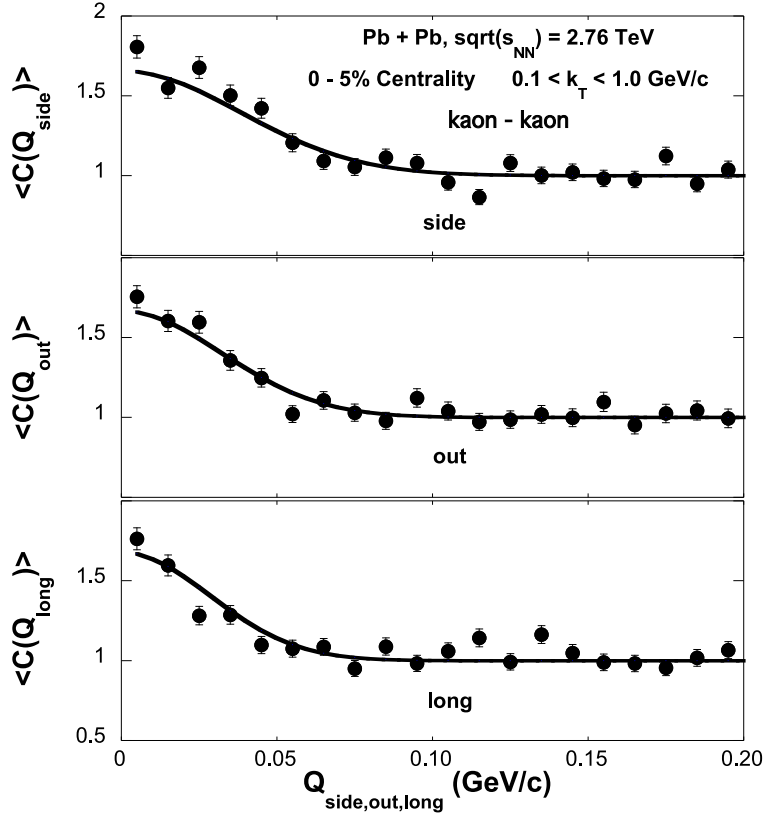} \caption{Sample two-kaon correlation
function with Gaussian fit projected onto the $Q_{out}$, $Q_{side}$, and $Q_{long}$
axes from the model. The collision centrality is $0-5\%$ and the $k_T$ bin is $0.1-1.0$ GeV/c.}
\label{fig8}
\end{center}
\end{figure}

\begin{figure}
\begin{center}
\includegraphics[width=140mm]{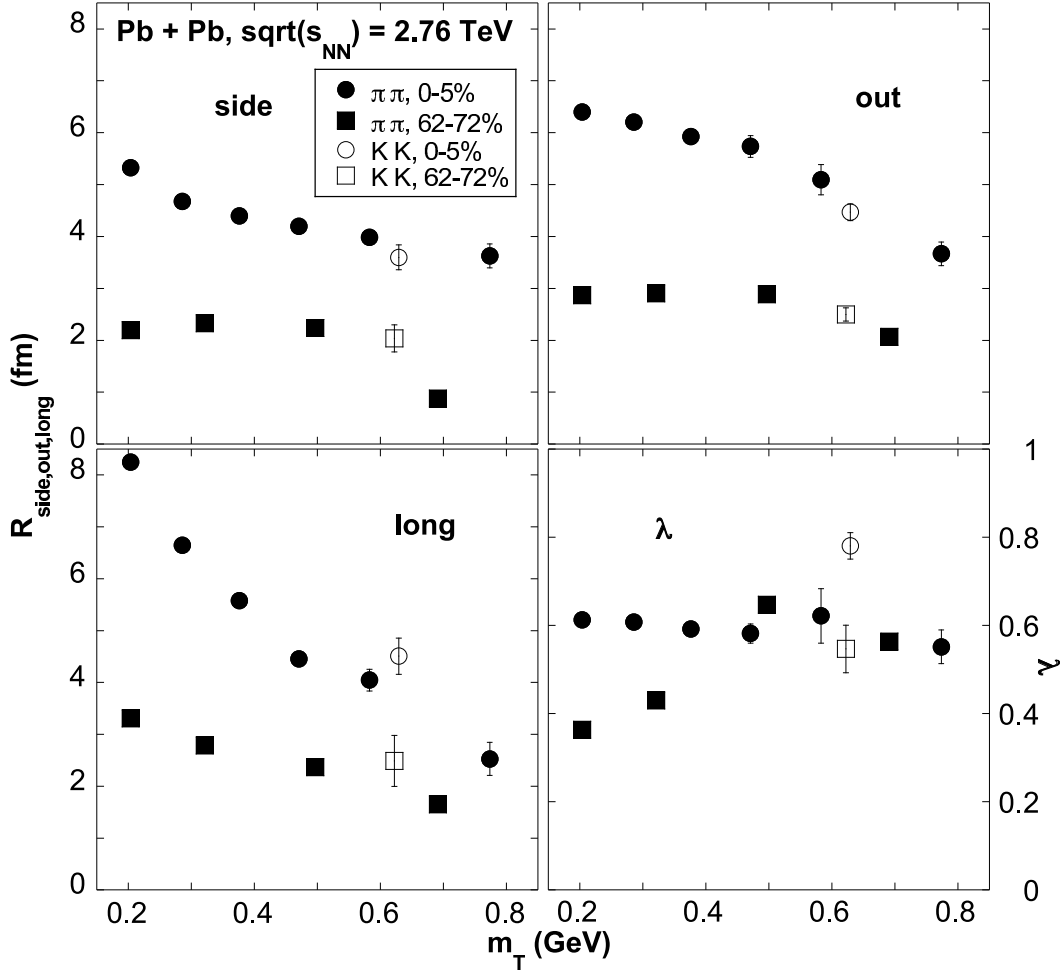} \caption{Comparison of model pion and kaon 
source parameters vs. $m_T$ 
for $Pb+Pb$ collisions, $-1<\eta<1$, $0-5\%$ and $62-72\%$ centralities.}
\label{fig9}
\end{center}
\end{figure}

Figures \ref{fig5}-\ref{fig9} and Tables \ref{table1}-\ref{table3} show predictions
from the model for two-pion and two-kaon HBT for 
$\sqrt{s_{NN}}=2.76$ TeV $Pb+Pb$ collisions.
For the HBT\cite{hbt1} calculations from the model, the three-dimensional 
two-boson correlation function is formed
and a Gaussian function in momentum difference variables is fitted to it to 
extract the boson source
parameters.  Boson statistics are introduced after the
rescattering has finished (i.e. when all particles have ``frozen out'')
using the standard method of pair-wise symmetrization of bosons in
a plane-wave approximation \cite{Humanic:1986a}. The three-dimensional
correlation function, $C(Q_{side},Q_{out},Q_{long})$, is then calculated 
in terms of the momentum-difference
variables $Q_{side}$, which points in
the direction of the sum of the two boson momenta in the transverse
plane, $Q_{out}$, which points perpendicular to $Q_{side}$ in the
transverse plane and the longitudinal variable along the beam
direction $Q_{long}$.

The final step in the calculation is extracting fit parameters by
fitting a Gaussian parameterization to the model-generated two-boson correlation 
function given by, \cite{Lisa:2005a}
\begin{eqnarray}
\label{e6}
\lefteqn{C(Q_{side},Q_{out},Q_{long}) = } \nonumber \\
& & G[ 1 + \lambda \exp( - Q_{side}^{2}R_{side}^{2} -
Q_{out}^{2}R_{out}^{2} - Q_{long}^{2}R_{long}^{2}) ]
\end{eqnarray}
where the $R$-parameters, called the radius parameters, are associated with each 
momentum-difference variable direction, G is a normalization constant, and
$\lambda$ is the usual empirical parameter added to help in the
fitting of Eq. (\ref{e6}) to the actual correlation function
($\lambda = 1$ in the ideal case). The fit is carried out in the conventional LCMS 
frame (longitudinally comoving system) in which the
longitudinal boson pair momentum vanishes \cite{Lisa:2005a}.
Figure \ref{fig5} shows a sample projected two-pion correlation function from the model with projected fit
to Eq. (\ref{e6}). It is seen that Eq. (\ref{e6}) fits the model-generated correlation function
quite well.

Figure \ref{fig6} shows the dependence of the model pion source parameters for
$Pb+Pb$ collisions on $k_T$ for centrality $0-5\%$ and $-1<\eta<1$
for the full calculation compared with the calculation with rescattering turned off.
For the full calculation, the radius parameters are all predicted to decrease with increasing $k_T$
showing the effects of ``flow''
as has been observed in RHIC $Au+Au$ collisions and 
elsewhere \cite{Lisa:2005a,Adams:2004yc}. The overall scales of $R_{side}$ and $R_{out}$
predicted for $Pb+Pb$ are comparable to those seen in RHIC $Au+Au$ collisions, whereas
$R_{long}$ is predicted to be about $25\%$ larger than at RHIC\cite{Adams:2004yc}.
The $\lambda$-parameter is seen to be mostly independent of $k_T$ with a value of about
0.6, which is significantly less than the ``ideal HBT case'' of $\lambda=1$. The main effect
causing $\lambda<1$ in the model is the presence of long-lived resonances such as
$\eta$ and $\eta'$ which decay into pions late in the collision thus suppressing 
the correlation function.
Looking at the radius parameters with rescattering turned off in Figure \ref{fig6}, it is seen that
their dependence on $k_T$ mostly disappears and their scales
are significantly reduced compared with the full calculation, showing the strong influence that rescattering has on the HBT parameters
in this model. Comparing $\lambda$-parameters, in the case with rescattering turned off the
$\lambda$-parameter is seen to increase with increasing $k_T$ unlike for the full calculation.
An explanation for this behavior is that with rescattering turned off, the pion source is less
Gaussian than with rescattering turned on and thus the $\lambda$-parameter tries to
compensate in the fit for this effect.

Figure \ref{fig7} compares the model pion source parameters vs. $k_T$ 
for $Pb+Pb$ collisions, where $-1<\eta<1$, for the two centralities $0-5\%$ and $62-72\%$.
Tables \ref{table1} and \ref{table2} give the values for the plots in Figure \ref{fig7}. The
more peripheral centrality case, i.e. $62-72\%$, is seen to have some qualitative similarities
with the ``Rescattering OFF'' calculation shown in Figure \ref{fig6}, in that the radius parameters
are seen to have weaker dependences on $k_T$ than for the more central case, i.e. $0-5\%$,
and are significantly smaller in magnitude. The $\lambda$-parameter $k_T$ dependence
for the peripheral case also resembles that for the ``Rescattering OFF'' case in that
$\lambda$ more or less increases with increasing $k_T$ as opposed to being mostly
independent of $k_T$ as for the full-calculation central case. The explanation for the similarities
between the peripheral case and the ``Rescattering OFF'' case is due to the smaller
particle multiplicity for the peripheral collisions and thus less rescattering present than for
the central case, determined by the smaller initial geometric overlap of the projectile-target
system in the peripheral case.

\begin{table*}
\caption{Model pion source parameters vs. $k_T$ 
for $\sqrt{s_{NN}}=2.76$ TeV $Pb+Pb$ collisions, $-1<\eta<1$, $0-5\%$ centrality.\label{table1}}
\begin{ruledtabular}
\begin{tabular}{cccccc}
$k_T$ range&$<k_T>$&$\lambda$&$R_{side}$&$R_{out}$&$R_{long}$ \\
(GeV/c)&(GeV/c)&   &(fm)&(fm)&(fm) \\ \hline
$0.10-0.20$&$0.15$&$0.613\pm0.000$&$5.33\pm0.00$&$6.40\pm0.00$&$8.25\pm0.00$ \\
$0.20-0.30$&$0.25$&$0.608\pm0.007$&$4.68\pm0.05$&$6.21\pm0.07$&$6.65\pm0.06$ \\
$0.30-0.40$&$0.35$&$0.592\pm0.003$&$4.40\pm0.07$&$5.93\pm0.11$&$5.58\pm0.05$ \\
$0.40-0.50$&$0.45$&$0.582\pm0.022$&$4.20\pm0.06$&$5.74\pm0.21$&$4.46\pm0.06$ \\
$0.50-0.70$&$0.57$&$0.622\pm0.062$&$3.99\pm0.13$&$5.10\pm0.29$&$4.05\pm0.21$ \\
$0.70-1.00$&$0.76$&$0.552\pm0.038$&$3.63\pm0.23$&$3.67\pm0.23$&$2.53\pm0.32$ \\
\end{tabular}
\end{ruledtabular}
\end{table*}

\begin{table*}
\caption{Model pion source parameters vs. $k_T$ 
for $\sqrt{s_{NN}}=2.76$ TeV $Pb+Pb$ collisions, $-1<\eta<1$, $62-72\%$ centrality.\label{table2}}
\begin{ruledtabular}
\begin{tabular}{cccccc}
$k_T$ range&$<k_T>$&$\lambda$&$R_{side}$&$R_{out}$&$R_{long}$ \\
(GeV/c)&(GeV/c)&   &(fm)&(fm)&(fm) \\ \hline
$0.10-0.20$&$0.15$&$0.364\pm0.003$&$2.20\pm0.02$&$2.87\pm0.01$&$3.31\pm0.02$ \\
$0.20-0.40$&$0.29$&$0.430\pm0.004$&$2.33\pm0.01$&$2.91\pm0.02$&$2.79\pm0.02$ \\
$0.40-0.60$&$0.48$&$0.647\pm0.009$&$2.24\pm0.06$&$2.89\pm0.05$&$2.37\pm0.07$ \\
$0.60-1.00$&$0.68$&$0.564\pm0.006$&$0.88\pm0.03$&$2.07\pm0.03$&$1.65\pm0.03$ \\
\end{tabular}
\end{ruledtabular}
\end{table*}

Model predictions have also been made for two-kaon HBT as shown in Figures \ref{fig8} and \ref{fig9}.
Figure \ref{fig8} shows a sample two-kaon correlation
function with Gaussian fit projected onto the $Q_{out}$, $Q_{side}$, and $Q_{long}$
axes from the model for the collision centrality  $0-5\%$, $-1<\eta<1$
and $k_T$ bin $0.1-1.0$ GeV/c. The Gaussian fit is seen to provide a reasonable fit
to the model correlation function. Figure \ref{fig9} presents a comparison of model pion and kaon 
source parameters vs. $m_T$ 
for $Pb+Pb$ collisions, $-1<\eta<1$, centralities $0-5\%$ and $62-72\%$. Table \ref{table3}
gives the values for the two-kaon source parameters plotted in Figure \ref{fig9}. A large $k_T$
bin of $0.10<k_T<1.00$ GeV/c with an average $k_T$ of
$<k_T>=0.39$ GeV/c was used for the two-kaon calculations in order to obtain reasonable statistical
errors from the 3200 event minimum bias $Pb+Pb$ run used in the present study. Also,
while a pseudorapidity range of $-1<\eta<1$ was used for the $0-5\%$ centrality
two-kaon calculation, a range of $-4<\eta<4$ was used for the $62-72\%$ centrality
two-kaon calculation in order to obtain reasonable statistical errors. As seen in Figure \ref{fig9},
the two-kaon calculations obey ``$m_T$-scaling'' reasonably well with the two-pion
calculations considering the large $k_T$ bin that was necessary to use for the kaons.

\begin{table*}
\caption{Model kaon source parameters vs. centrality bin
for $\sqrt{s_{NN}}=2.76$ TeV $Pb+Pb$ collisions, $0.10<k_T<1.00$ GeV/c
 ($<k_T>=0.39$ GeV/c).\label{table3}}
\begin{ruledtabular}
\begin{tabular}{ccccc}
Centrality bin&$\lambda$&$R_{side}$&$R_{out}$&$R_{long}$ \\
   &   &(fm)&(fm)&(fm) \\ \hline
$0-5\%$&$0.781\pm0.030$&$3.60\pm0.24$&$4.47\pm0.16$&$4.51\pm0.35$ \\
$62-72\%$&$0.547\pm0.054$&$2.04\pm0.26$&$2.50\pm0.13$&$2.49\pm0.49$ \\
\end{tabular}
\end{ruledtabular}
\end{table*}

\section{Summary and Conclusions\label{conclusions}}
A kinematic model based on the superposition of PYTHIA-generated $p+p$ collisions, relativistic geometry and final-state hadronic rescattering
has been used in the present work to predict several hadronic observables in $\sqrt {s_{NN}} = 2.76$ TeV $Pb+Pb$ collisions. A short proper time for hadronization of $\tau=0.1$ fm/c
has been assumed as in previous studies with this model which have shown qualitative
agreement with experiments.
Predictions for rapidity and transverse momentum distributions, elliptic flow, and two-boson femtoscopy have been presented which will likely be among the first observables to be extracted from analyses
of the first $Pb+Pb$ data from the LHC.

The most noticeable features
of the predictions from the present model study which have been presented are summarized below:
\begin{itemize}
\item $dn/d\eta$ near mid-rapidity for charged particles in LHC $Pb+Pb$ collisions is predicted to be about 1200 for
a $0-5\%$ centrality window. This puts its value at the lower end of the range of predictions which have been recently made of $1500-4000$ in central collisions using various extrapolations of RHIC experimental rapidity densities.
\item The $0-5\%$ centrality charged particle $p_T$ distribution for $p_T>5$ GeV/c is predicted to be about two orders of magnitude larger at the LHC compared with RHIC.
\item High-$p_T$ suppression of the $R_{AA}$ is predicted to be present in LHC $Pb+Pb$
collisions as has been seen in RHIC collisions.
\item Elliptic flow for charged hadrons is predicted to be comparable to that seen in RHIC collisions.
\item Two-pion HBT radius parameters from LHC $Pb+Pb$ are predicted 
to be comparable in scale to those from RHIC
$Au+Au$ collisions for $R_{side}$ and $R_{out}$, and ~25\% larger for $R_{long}$ and are
predicted to
show decreasing magnitude with increasing $k_T$, i.e. ``flow'' effects. Also, two-kaon HBT radius parameters
are predicted to show ``$m_T$-scaling.''
\end{itemize}

As mentioned earlier, these are ``limiting case scenario'' predictions from a model that
only considers hadronic degrees of freedom, i.e. which assumes that only ``ordinary'' physics 
processes are taking place in these collisions. Of course the ``best case scenario'' will be that
these predictions disagree wildly with 
the actual measured hadronic observables which will soon be extracted 
by LHC experiments and exotic phenomena will indeed be observed.



\begin{acknowledgments}
The author wishes to acknowledge financial support from the U.S.
National Science Foundation under grant PHY-0970048, and to acknowledge computing
support from the Ohio Supercomputing Center.
\end{acknowledgments}

\end{document}